\documentclass[runningheads]{cl2emult}

\usepackage{makeidx}  
\usepackage{graphicx} 
\usepackage{subeqnar} 
\usepackage{multicol} 
\usepackage{cropmark} 
\usepackage{eso}      
\makeindex            

\def\simlt{\lower.5ex\hbox{$\; \buildrel < \over \sim \;$}}
\def\simgt{\lower.5ex\hbox{$\; \buildrel > \over \sim \;$}}
  
\def\eg{{\rm e.g.$\;$}}

\def\etal{{\rm et al.$\;$}}

\def\ds{\displaystyle}
\def\l#1{\left#1}
\def\r#1{\right#1}

\def\maxima{{\sc maxima-i}}
\def\boom{{\sc boomerang-ldb}}

\def\iNct{N_{Ct}^{-1}}

\begin{document}
\title*{From the time-ordered data to the Maximum-\protect\newline Likelihood
temperature maps of the Cosmic Microwave Background anisotropy.}
\titlerunning{Maximum-Likelihood CMB maps}

\author{Radek Stompor\inst{1,2,3}
\and Amedeo Balbi\inst{4}
\and Julian Borrill\inst{5,1}
\and Pedro Ferreira\inst{6,7}
\and Shaul Hanany\inst{8,1}
\and Andrew Jaffe\inst{1,9,10}
\and Adrian Lee\inst{10,1,11}
\and Sang Oh\inst{1,10}
\and Bahman Rabii\inst{1,10}
\and Paul Richards\inst{1,10}
\and George Smoot\inst{1,11,2}
\and Celeste Winant\inst{1,10}
\and Jiun-Huei Proty Wu\inst{9}
}
\authorrunning{Stompor et al.}

\institute{Center for Particle Astrophysics, University of California, Berkeley, CA 94720, USA
\and Space Sciences Laboratory, University of California, Berkeley, CA 94720, USA
\and Copernicus Astronomical Center, Warszawa, Poland
\and Dipartimento di Fisica, Universit\`a Tor Vergata, Roma, Italy
\and NERSC, Lawrence Berkeley National Laboratory, Berkeley, CA 94720, USA
\and Astrophysics, University of Oxford, NAPL, Oxford, OX1 3RH, UK
\and CENTRA, Instituto Superior Tecnico, Lisboa, Codex, Portugal
\and School of Physics and Astronomy, University of Minnesota, Minneapolis, MN55455, USA
\and Dept. of Astronomy, University of California, Berkeley, CA 94720, USA
\and Dept. of Physics, University of California, Berkeley, CA 94720, USA
\and Lawrence Berkeley National Laboratory, Berkeley, CA 94720, USA
}

\maketitle              

\begin{abstract}
We review selected methods of the Cosmic Microwave Background data
analysis appropriate for the analysis of the largest currently available data sets.
We focus on techniques of the time-ordered data manipulation and map making algorithms based on the maximum-likelihood approach. 
The presented methods have been applied to the \maxima\ data analysis
(Hanany \etal 2000) and the description of the algorithms is 
illustrated with the examples drawn from that experience.
The more extensive presentation  of the here-mentioned issues will be given in the forthcoming paper (Stompor \etal 2001).
\end{abstract}

\section{Introduction}
The growing size and complexity of the available CMB data sets poses a difficult challenge for the data analysis (\eg Borrill (1999), 
Borrill \etal (2000)). The largest
already existing data sets (\eg \boom, \maxima) consist of up to many million of the time measurements sampling areas of 
the sky of up to many thousands of the beam-size pixels. The analysis of such a big data set is 
usually divided into three distinct stages focusing in turn on the time-ordered data, maps and power spectra.
On each stage the size of the data item to be manipulated is significantly reduced.
However, a successful implementation of such a program
requires that the compression performed on each stage should 
should not compromise the cosmological information contained in the data. 
The algorithms presented in the following are designed to be optimal
-- or nearly optimal -- in a maximum-likelihood sense 
under the assumptions of the Gaussianity and stationarity of the instrumental noise.

We start from the description of the time-ordered data manipulation techniques which prove to be necessary to ensure the efficiency and
applicability of the map making algorithms. Then we briefly review a variety of map making algorithms
highlighting involved assumptions and assessing their efficiency.

\section{Noise estimation and gaps filling}

In the usual map-making methodology the noise power spectra are taken to be given with an arbitrary high precision. The circumstances
in which actual CMB data are collected are specific enough to require the
noise estimation to be evaluated directly from
the time-ordered data. The robust noise-estimation procedure therefore
needs to be capable of efficient dealing with the undesirable effects present in
the real data.

The common feature of the realistic time streams are short breaks (gaps) in their continuity caused \eg by the cosmic
rays or other transients in the experimental apparatus. 
Their presence not only prevents a straightforward application of the Fast Fourier techniques (FFT)
to the noise estimation but also introduces a whole suite of subtle problems related to 
their treatment on the level of the time-ordered data manipulation. They also appear to be one of
major obstacles in an implementation of the efficient and fast map making algorithms. 
Therefore restoration of the continuity of the time-ordered data -- in a statistically correct way -- stands out as one of the important 
goals of the data analysis on the time domain stage. A required algorithm needs simultaneously to estimate the noise
power spectrum and fill the gaps with pure Gaussian noise realization without compromising its correlations throughout the 
length of the time stream. As described in detail in Stompor \etal (2001) that can be achieved iteratively when on each step of the iterations
the gaps are filled with the constrained realization (Hoffman \& Ribak 1991) of the Gaussian noise, a power spectrum of which has been determined
on the preceding step of the iterations using standard FFT methods. 

The noise estimation procedure
attempts to estimate a noise ensemble average power spectrum using
just one realization of the time stream. That is clearly not sufficient.
As a result of the above sketched algorithm one usually ends up with the reliable estimation of the
ensemble average noise power spectrum in time domain at sufficiently
high frequencies (for \maxima\ for $f\simgt 0.1$Hz), but the low
frequency end is a subject to a non-negligible sampling
error. To minimize the effect of the low frequencies we marginalize over the low frequency
part of the spectrum. No significant dependence on the assumed
low frequency cut-off (in the range from $0.01\;$Hz up to $0.4\;$Hz) has been found in 
the \maxima\ case.

Adding randomly drawn signal to the data introduces an extra freedom and may undermine the uniqueness of the
results. Though depending on the particular noise realization the results may somewhat differ,
all of them are statistically equivalent by construction, and 
no bias is introduced. Moreover, that extra randomness 
can be minimized by introducing a fictitious gap pixel to the recovered map.
As a result we find that the final maps and their power spectra for
\maxima\ are robust and not affected by that procedure.

So far, for simplicity we have been assuming that the time-ordered
data are noise dominated, however, an extension of the described
algorithm for the more general case is straightforward (Ferreira \& Jaffe 2000).

\section{Map making}

The general algebra for the maximum-likelihood map making under the assumption of the
Gaussian correlated noise can be found in \eg Tegmark (1997). In short, if the time
stream data $d$ can be modeled as
\begin{equation}
d=Am+n, \label{simpletimestream}
\end{equation}
where $d$ is the vector of the measurements for the given chunk of the time
stream, $A$ a pointing matrix assigning each of the time samples to an
appropriate pixel in the sky and 
$n$ - a time stream noise - is a vector of the random Gaussian variables with 
correlations as given by $N_t$ - a time domain correlation matrix,
then a (maximum-likelihood) map $m$ and its noise-correlation matrix in
pixel domain ($N_p$) are given by,
\begin{eqnarray}
m&=&\l(ADMD^TA^T\r)^{-1}\,ADMd\label{mapmaking}\\
N_p&=&\l(ADMD^TA^T\r)^{-1}\l(ADMN_tMD^TA^T\r)\l(ADMD^TA^T\r)^{-1}
\end{eqnarray}
here $M$ is a positive definite symmetric matrix. 
$D$ is a time domain representation of the filters applied to the time
stream on the previous stages (\eg a prewhitening filter) assumed to
be orthogonal $DD^T=1$.

If $M=N_t^{-1}$ then $m$ is a minimum variance
map. The other choices can sacrifice the optimality
but being computationally faster. In particular Tegmark (1997)
proposed to choose as $M$ only a circulant part of the $N_t$ matrix.
In the following we discuss a number of approaches and their
applications to a realistic \maxima-like data set.

\subsection{Minimum variance variants}

\subsubsection{Exact implementation.}
The implementation of the exact minimum variance estimator of the map
seems a daunting task. The time stream length ($n_t$) may easily reach many millions
of the time samples making an inversion of the time domain noise
correlation matrix -- an ``$n_t$-cubed'' process -- prohibitive. However,
for a \maxima\ like experiment with the time stream chunks of the length reaching
``only'' up to $2.5\times 10^5$ time samples we found that the 
implementation is feasible even on a moderately large workstation. 

At the heart of that implementation lies a simple observation that the time
domain correlation matrix of the stationary and continues time stream 
is Toeplitz \eg $N_{t}\l(i,j\r)=N_{t}\l(i-j,0\r)$. The inversion
of the the Toeplitz matrix can be performed in as few as $n_t^2$ operations. A
clearly feasible task for $\sim 10^5$ time samples. (An even faster
algorithm exists (\eg Golub \& van Loan 1983) bringing that number down to $\sim
n_t\log^2 n_t$.) An extra gain in a number of operations can be achieved
if the noise-correlation length is shorter than the time stream 
length and the noise-correlation matrix band-diagonal.

The actual time streams are
commonly strewn with the gaps and a Toeplitz character of
the correlation matrix lost. However the gap filling
algorithm, presented earlier, reconstructs the continuity of the
time stream and its stationarity. To minimize the entirely spurious
content of the gaps being added to the map all the time
samples from the gaps are directed to an extra fictitious pixel (``a
gap pixel'') which is subsequently rejected (marginalized over) from
the map and the pixel domain noise-correlation matrix. 

The computational effort for the case when $D\equiv 1$
scale with the number of pixels $n_{pix}$ and the number of
time samples $n_t$ as:  
\begin{itemize}
\item{noise inverse in time domain: $N_t \longrightarrow N_t^{-1}$ : $\propto n_t^2$;}
\item{noise inverse in pixel domain: $N_t^{-1}, A \longrightarrow A N_t^{-1} A^T$ : $\propto n_t^2$;}
\item{noise weighted map: $N_t^{-1}, A, d \longrightarrow A N_t^{-1} d$ : $\propto n_t^2$;}
\item{noise matrix in pixel domain: $N_p \longrightarrow N_p^{-1}$ : $\propto n_{pix}^3$;}
\item{final map: $N_p^{-1}, A N_t^{-1} d \longrightarrow N_p^{-1} A N_t^{-1} d $ $\propto n_{pix}^2$.}
\end{itemize}
In the case of the first three items from the list the substantial savings can be made if the
fact that the inverse noise-correlation matrix is assumed to be band-diagonal -- an approximation 
usually well fulfilled for an inverse of a band-diagonal noise-correlation matrix.
If the noise-correlation matrix is sparse then the additional savings can also be made.

Memory-wise  clearly there is only need to keep a first row of the $N_t$ matrix 
and in a typical situation the major requirement would be then set by the size of the
$AN_t^{-1}$ matrix which is $\propto n_{pix}n_t$ and the size of the
noise correlation for the entire map $\propto n_{pix}^2$.
Note that the first of these limits can be alleviated at the expense
of the operation counts (there is no need to keep all $n_{pix} n_t$
matrix at the same time but then some of the parts may have to be
recomputed a multiple number of times).

\subsubsection{Approximate approach.}

The way to speed up map making but still aiming at the optimal minimum
variance map was proposed by Ferreira \& Jaffe (2000).
The approximation those authors favor assumes that,
\begin{equation}
N_{t}^{-1}\l(i,j\r)\simeq \l\{
\begin{array}{l}
{\ds N_{Ct}^{-1}\l(i,j\r) \mbox{\ \ \ \rm if\ \ }|i-j| \le
\min\l(n_c/2,l_c\r) }\\
\\
{\ds 0, \mbox{\ \ \ \ \ \ \ \rm otherwise.} }
\end{array}
\r.
\label{FJapprox}
\end{equation}
Here $n_c$ and $l_c$ are the time chunk and correlation lengths
respectively, and a subscript $C$ denotes a circulant part of the noise
correlation matrix defined as $N_C\l(i,j\r) = N_C\l(i-j,0\r) =
N_C\l(n_c-i,0\r)$ for $0 \le i\le n_c/2$.

This approach is designed to perform the $N_t$ inversion with the ``FFT'' speed 
providing at the same time a good approximation to the exact
solution of the previous section. By construction, however, it can be only exact in the absence of 
the noise correlations.

The operation count changes only for two, but the most time consuming, items from the list in the
previous paragraph which now read,
\begin{itemize}
\item{noise inverse in time domain: $N_t \longrightarrow N_t^{-1}$ : $\propto n_t\log n_t$.}
\item{noise weighted map: $N_t^{-1}, A, t \longrightarrow A N_t^{-1} t$ : $\propto n_t \log n_t$;}
\end{itemize}
The scaling of the noise
weighted map computation is given assuming using FFT and a Toeplitz
matrix property (Golub \& van Loan 1983).

The required memory is set by the size of the noise matrix in pixel
domain $\propto n_{pix}^2$. It is apparent that if the efficient fast
(``$n_t\log^2 n_t$'') implementation of the Toeplitz matrix inversion is
available then the major computational advantage of the approximate method over
the exact one would vanish and only the memory requirement would
remain as its main asset.

This approximation is implemented in the very successful, publicly
available package called MADCAP by Julian Borrill (\eg Borrill \etal 2000).

\subsection{Circulant variants}

In this case $M=N_{Ct}^{-1}$ and the noise-correlation matrix in pixel domain can be
expressed as follows (Tegmark 1997),
\begin{eqnarray}
N_p&\equiv& \l(AD\iNct D^TA^T\r)^{-1}+\label{circnoise1}\\
&+&\l(AD\iNct D^TA^T\r)^{-1}\l(AD\iNct N_{S t}\iNct
D^TA^T\r)\l(AD\iNct D^TA^T\r)^{-1},\nonumber
\end{eqnarray}
where we decomposed explicitly a noise-correlation matrix in time
domain into its circulant ($N_{C t}$) and sparse parts ($N_{S t}$),
$N_t\equiv N_{C t}+N_{S t}$.  
The sparse part of the matrix corrects for the presence of 
``corners'' of the circulant matrix and the gaps in the time stream.
As advocated by Tegmark (1997) such a matrix should be
really sparse and a number of the required operation greatly decreased over the
exact approach described above. In the implementation of that author
the $D$ filters are additionally used to enhance sparseness.

The price to pay for it is not only a non-optimal map as a result (the loss of the precision is hoped not to be 
substantial here), but also definitely
more complicated algebra to be implemented especially in a presence of
large number of gaps in the time stream. If no assumption about the
band-diagonality is made then the computational costs scale as follows:
\begin{itemize}
\item{noise inverse in time domain: $N_t \longrightarrow N_t^{-1}$ : $\propto n_t\log n_t$;}
\item{noise inverse in pixel domain: 
\begin{description}
\item{circulant part:}{ $N_t^{-1}, A \longrightarrow A N_t^{-1} A^T$ : $\propto n_t^2$;}
\item{sparse part:}{ $N_{S t},M,A \longrightarrow AMN_{S t}MA^T$ : $\propto n_{pix}n_t^2$;}
\end{description}
}
\item{noise weighted map: $N_t^{-1}, A, d \longrightarrow A N_t^{-1} d$ : $\propto n_t \log n_t$;}
\item{noise matrix in pixel domain: $N_p \longrightarrow N_p^{-1}$ : $\propto n_{pix}^3$.}
\end{itemize}

We have omitted the filter matrix $D$ in above expressions assuming
that both $M$ and $D$ are circulant and therefore their product can be
computed with the ``FFT speed''.
The effort is dominated usually by the sparse part
computation ($n_t > n_{pix}$) though the above scaling includes only
the fastest growing term which can not be suppressed if only the band
diagonality of the $N_t$ matrix is assumed.
Memory required for performing that task is $\propto n_{pix}n_t$.

Clearly if no more approximations are made this
approach is far from competitive with any of those discussed above.
One possible way of improving on that is to recognize the fact that if
$N_{C t}$ is band-diagonal (plus of course non-zero corners to fulfill
the circulancy criterion) then also 
$N_{C t}^{-1}$, $D^TN_{C t}^{-1}$ and $D^TN_{C t}^{-1}D$ are usually approximately band-diagonal.

Another possible approximation, referred to hereafter, is to
neglect completely the sparse correction in the expression for the noise-correlation matrix in the pixel domain.
Clearly the operational cost goes down
to $n_t^2$ or $n_{pix}^3$ -- whichever larger -- and is usually
$\propto n_{pix}^3$ especially if a band diagonal character of the inverse of
$N_{C t}^{ -1}$ is assumed. In such a case the method achieves the
performance 
comparable with that of
Ferreira \& Jaffe (2000).

\subsection{Comparison and assessment}

In the paper by Stompor \etal (2001) we show that though none of discussed above 
approximations reproduces perfectly the map and the noise-correlation
matrix in the pixel domain ($N_{p}$), it preserves the statistical
information contained in the data -- \eg 2-point correlation
properties -- correctly.
In fact we find that both the approximations, as described above,
fare very well in recovering the anisotropy power spectrum 
with no apparent systematic bias.  

Clearly the performance of these approximations needs to 
be further tested if other statistics are to be applied to the maps.

Out of the two exact methods the minimum variance implementation described here performs generally much better.
The very fact that the exact maximum-likelihood method can be applied 
to the data sets of the \maxima\ size is of importance for the further development, tests and validation
of the approximate algorithms.
The simple operation counts and memory requirements are clearly great
assets of those  methods and major stumbling blocks for the exact approach
to go far beyond the size of the current data sets. 

It is worth noticing that the gaps filling procedure described in the beginning is
important for all the described methods: in a case of 
the exact methods (optimal and circulant) it facilitates the practical feasibility of the 
implementation.
For the approximate methods it improves the accuracy and robustness of the involved 
approximations.

\section{Summary}

We briefly have discussed here the making of the CMB temperature anisotropy maps from the time-ordered data.
We have argued that for the data set with up to a few millions of time samples producing the maps 
of up to a few thousands of the pixels is readily possible without
need for sacrificing the optimal character of the final map. Moreover
accurate and efficient approximate methods also exist. Those have been
tested through
the direct comparison with the exact method on the realistic
simulations and the \maxima\ data set and can become a starting point
for developing algorithms capable of coping with 
still bigger and more complicated data sets. It seems likely that the hard problem of this kind
of data analysis will be accounting on all kinds of imperfectness of a real data set 
in the statistically sound way and without the substantial
increase of the operational count, rather than solely a sheer size of a data set itself. 
As an example here we have demonstrated how to incorporate the presence of short discontinuities in the time-ordered data.
The other common problems are to be discussed in Stompor \etal (2001).


RS acknowledges support of NASA Grant NAG5-3941 and help of Polish
State Committee for Scientific Research grant no.$\;$ 2P03D01719.
AHJ and JHPW acknowledge support from 
NASA LTSA Grant no.$\;$ NAG5-6552 and NSF KDI Grant no.$\;$ 9872979. 
PGF acknowledges support from the RS. 
BR and CDW acknowledge support from NASA GSRP
Grants no.$\;$ S00-GSRP-032 and S00-GSRP-031.

\clearpage
\addcontentsline{toc}{section}{Index}
\flushbottom
\printindex


\begin{thebibliography}{7}
%
\addcontentsline{toc}{section}{References}

\bibitem{ref1.1} Hanany, S., \etal, (2000), \maxima: A Measurement of the Cosmic Microwave Background Anisotropy 
on angular scales of 10 arcminutes to 5 degrees. {\it Astrophys. J. Letters}, in press

\bibitem{ref1.2}Stompor, R., \etal, (2001), {\it in preparation}

\bibitem{ref1.3} Borrill, J., (1999) in EC-TMR Conference Proceedings,
476, 3K Cosmology, ed. L. Maiani, F. Melchiorri, \&\ N. Vittorio
(Woodbury, New York: AIP), 224

\bibitem{ref1.4} Borrill, J., Ferreira, P.,G., Jaffe, A.,H., \&\ Stompor, R. (2000), CMB Data Analysis with MADCAP, this volume

\bibitem{ref1.5} Hoffman, Y. \&\ Ribak, E., (1991), {\it Astroph. J. Letters}, {\bf 380}, 5

\bibitem{ref1.6} Tegmark, M., (1997), CMB mapping experiments,
Phys. Rev. D, {\bf 56}, 4516

\bibitem{ref1.7} Golub, H., \&\ van Loan, F., (1983), Matrix
operations, The Johns Hopkins University Press

\bibitem{ref1.8} Ferreira, P., G., \&\ Jaffe, A.,H., (2000), Simultaneous estimation of noise and signal 
in cosmic microwave background experiments, M.N.R.A.S., {\bf 312}, 89

\end{thebibliography}
\end{document}